\documentclass[twocolumn,pra,showpacs]{revtex4}
\usepackage{amssymb}
\usepackage{amsfonts}
\usepackage{amsmath}
\usepackage{graphicx}

\begin{document}

\title{Symmetries of general non-Markovian Gaussian diffusive unravelings}
\author{Adri\'{a}n A. Budini}
\affiliation{Consejo Nacional de Investigaciones Cient\'{\i}ficas y T\'{e}cnicas
(CONICET), Centro At\'{o}mico Bariloche, Avenida E. Bustillo Km 9.5, (8400)
Bariloche, Argentina, and Universidad Tecnol\'{o}gica Nacional (UTN-FRBA),
Fanny Newbery 111, (8400) Bariloche, Argentina}
\date{\today }

\begin{abstract}
By using a condition of average trace preservation we re-derive a general
class of non-Markovian Gaussian diffusive unravelings [L. Di\'{o}si and L.
Ferialdi, Phys. Rev. Lett. \textbf{113}, 200403 (2014)], here valid for
arbitrary non-Hermitian system operators and noise correlations. The
conditions under which the generalized stochastic Schr\"{o}dinger equation
has the same symmetry properties (invariance under unitary changes of
operator base) than a microscopic system-bath Hamiltonian dynamics are
determined. While the standard quantum diffusion model (standard noise
correlations) always share the same invariance symmetry, the generalized
stochastic dynamics can be mapped with an arbitrary bosonic environment only
if some specific correlation constraints are fulfilled. These features are
analyzed for different non-Markovian unravelings equivalent in average.
Results based on quantum measurement theory that lead to specific cases of
the generalized dynamics [J. Gambetta and H. M. Wiseman, Phys. Rev. A 
\textbf{66}, 012108 (2002)]\ are studied from the perspective of the present
analysis.
\end{abstract}

\maketitle


\section{Introduction}

The theory of open quantum systems is well established when a Markovian
approximation applies \cite{breuerBook}. In this situation the density
matrix evolution is defined by a Kossakovski-Lindblad equation. In addition,
the system dynamics can be read in terms of an ensemble of stochastic
trajectories developing in the system Hilbert space \cite%
{barchiellibook,breuerBook,carmichel,plenio,milburnbook,gisin,rigo,howard,ghirardi}%
. Different stochastic Schr\"{o}dinger equations characterize the ensemble
dynamics (unravelings). The diffusive case \cite{barchiellibook} corresponds
to multiplicative Gaussian white noises. It allows to describe quantum
systems subjected to a continuous measurement process \cite%
{carmichel,plenio,milburnbook}\ as well as to formulate dynamical wave
vector collapse models \cite{ghirardi}.

In the last years, with the goal of establishing a non-Markovian extension
of the standard Markovian open quantum system theory, different research
lines were opened. In particular, the stochastic ensemble representation of
a quantum system coupled to a bosonic bath \cite{diosi1,diosi2} triggered
the study of stochastic Schr\"{o}dinger equations driven by multiplicative 
\textit{non-white} Gaussian noises. As demonstrated in the seminal
contributions of Strunz and Di\'{o}si \cite{diosi1}, the emerging
time-evolution-equation involves a functional derivative of the wave vector.
Due to this feature, the formulation of different derivations and
perturbation schemes, which lead to evolutions without involving an
undetermined functional derivative, is a topic that is of interest up to
present time \cite{yu,cresser,budini, gamba,lilam,jerarquia,wu}. On the
other hand, exact closed expressions for the functional derivative were
found in different physical situations \cite%
{strunz,yuyu,jing,eplYu,feria,YuBipartite}.

The non-Markovian quantum diffusion model \cite{diosi1,diosi2} provided an
alternative context for discussing non-Markovian continuous measurement
theory \cite{gambetta,hidden,Diosol}, non-Markovian extensions of
spontaneous wave function collapse models \cite{reduction,bassilo}, and
operator correlations beyond the quantum regression theorem \cite{alonsoQRT}%
. The formalism was also used to characterize specific physical systems such
as for example quantum Brownian motion \cite{strunzBrownian}, complex
optical arranges \cite{eisfeld,suess}, charge transport in organic crystals 
\cite{chinos} and many body systems \cite{you}.

A relevant advance in the field was introduced recently by Di\'{o}si and
Ferialdi \cite{DiosiPRL}. On the basis of a path integral approach, they
defined a generalized class of non-Markovian Gaussian stochastic Schr\"{o}%
dinger equations. Similarly to the Markovian case, the generalized
unraveling is \textit{parametrized} by a set of complex noise correlations
that do not affect the density matrix evolution \cite{rigo,howard}. For
Markovian unravelings the degree of freedom introduced by the extra
correlations can be set from symmetries constraints such as the invariance
of the unraveling under linear system operator transformations under which
the corresponding master equation is invariant \cite{gisin,rigo,howard}. The
main goal of this paper is to develop similar symmetry analysis for the
generalized non-Markovian unraveling \cite{DiosiPRL} and to find which
constraints on the noise correlations arise.

We show that both the generalized non-Markovian Schr\"{o}dinger equation and
its associated density matrix evolution \textit{always} share the same
symmetry property, that is, they are invariant under arbitrary unitary
changes of the system operator base. Therefore, in contrast to the Markovian
case, we focus on which constraints on the noise correlations may arise when 
\textit{mapping} the unravelling invariance with that of a microscopic
system-bath Hamiltonian description. Different solutions to this problem are
found, which are based on different kind of mapping between the noise and
bath operator correlations. The emerging conditions under which the mapping
is consistent are studied for a single dissipative channel dynamics.
Furthermore, the stretched relation of the generalized non-Markovian
unraveling with previous results derived from quantum-measurement theory 
\cite{gambetta} are analyzed from this perspective.

We base the present analysis on a generalized non-Markovian Gaussian Schr%
\"{o}dinger equation that, in contrast to Ref. \cite{DiosiPRL}, become
written in an arbitrary non-Hermitian base of system coupling operators.
Instead of a path integral formalism, here the wave vector evolution [Eq. (%
\ref{General})] is derived by postulating a stochastic density matrix
dynamics driven by multiplicative non-white Gaussian noises, where an
undetermined contribution is obtained from a condition of average trace
conservation~\cite{budini}.

The paper is organized as follows. In Sec. II, based on the stochastic
approach outlined previously, we obtain the general non-Markovian Gaussian
diffusive unraveling. The main results are obtained in Sec. III, where its
symmetry properties are analyzed. In Sec. IV we exemplify the main results
by studying the case of a single Hermitian coupling channel. Previous
results obtained from quantum measurement theory are also studied and
derived by using the present approach. The Conclusions are given in Sec. V.
In the Appendix we define the main properties of the noise correlations as
well as a generalization of Novikov theorem \cite{novikov}\ valid for
arbitrary multiplicative complex Gaussian noises.

\section{General stochastic Gaussian dynamics}

Both in the Markovian and non-Markovian regimes, stochastic Schr\"{o}dinger
equations define an ensemble of system states, which can be written in terms
of a stochastic density matrix $\rho _{st}(t).$\ Its average over the
ensemble of realizations, denoted as $\langle \cdots \rangle ,$\ gives the
system state, $\rho (t)\equiv \left\langle \rho _{st}(t)\right\rangle .$ In
the present approach, we start by postulating the time evolution of $\rho
_{st}(t),$%
\begin{eqnarray}
\frac{d}{dt}\rho _{st}(t) &=&-i[H_{S},\rho _{st}(t)]-\mathcal{U}[\rho _{st}]
\notag \\
&&-i\lambda (\mathcal{F}(t)\rho _{st}(t)-\rho _{st}(t)\mathcal{F}^{\dagger
}(t)).  \label{RhoStochastic}
\end{eqnarray}%
Here, $H_{S}$ is the system Hamiltonian. $\lambda $ is coupling parameter.
The fluctuation operator reads%
\begin{equation}
\mathcal{F}\left( t\right) \equiv \sum\limits_{\alpha }z_{\alpha
}(t)L_{\alpha },  \label{efe}
\end{equation}%
where $\{L_{\alpha }\}$ is the set of coupling system operators. The set of
(multiplicative) complex noises $\{z_{\alpha }(t)\}$ are Gaussian with null
mean value $\langle z_{\alpha }\left( t\right) \rangle =0,$ and correlations%
\begin{equation}
\chi _{\alpha \beta }(t,s)\equiv \langle z_{\alpha }^{\ast }\left( t\right)
z_{\beta }\left( s\right) \rangle ,  \label{corre1}
\end{equation}%
and also%
\begin{equation}
\eta _{\alpha \beta }(t,s)\equiv \langle z_{\alpha }\left( t\right) z_{\beta
}\left( s\right) \rangle .  \label{corre2}
\end{equation}%
Notice that in contrast with previous analysis \cite{budini} and similarly
to Ref. \cite{DiosiPRL}, here we are taking into account the correlations $%
\eta _{\alpha \beta }(t,s).$ These objects allow to consider arbitrary
correlated noises (see Appendix). While the following derivation is similar
to that when $\eta _{\alpha \beta }(t,s)\rightarrow ~0$ \cite{budini}, here
we show it in order to enlighten the main assumptions over which it relies.

The unknown functional $\mathcal{U}[\rho _{st}]$ is determined by average
trace condition%
\begin{equation}
\frac{d}{dt}\mathrm{Tr}\langle \rho _{st}(t)\rangle =0,
\label{TraceConserva}
\end{equation}%
and the separability condition (pure state unraveling)%
\begin{equation}
\rho _{st}(t)=\left\vert \psi \left( t\right) \right\rangle \left\langle
\psi \left( t\right) \right\vert .  \label{separar}
\end{equation}%
This last requirement allow us to defining a stochastic Schr\"{o}dinger
equation for the system state $\left\vert \psi \left( t\right) \right\rangle
.$ In addition, this requisite implies that the stochastic map $\rho
_{st}(0)\rightarrow \rho _{st}(t)$ is a completely positive one \cite%
{breuerBook}. Therefore, Eq. (\ref{separar}) is a sufficient condition that
guarantees the completely positive property of the density matrix evolution $%
\rho (0)\rightarrow \rho (t).$

By averaging Eq. (\ref{RhoStochastic}) and imposing condition (\ref%
{TraceConserva}) we get%
\begin{equation}
\mathrm{Tr}\left\langle \mathcal{U}[\rho _{st}]\right\rangle =-i\lambda 
\mathrm{Tr}\left( \left\langle \mathcal{F}\left( t\right) \rho
_{st}(t)\right\rangle -\left\langle \rho _{st}(t)\mathcal{F}^{\dagger
}\left( t\right) \right\rangle \right) .  \label{trazon}
\end{equation}%
Novikov theorem \cite{novikov} gives an exact (functional) expression for
the average of a product between a Gaussian noise and an arbitrary
functional of it. In the Appendix we provide its generalization for
arbitrary complex noises. Hence, from Eqs. (\ref{efe}) and (\ref{Novikov})
the previous contributions can be written as%
\begin{eqnarray}
\langle \mathcal{F}\left( t\right) \rho _{st}(t)\rangle
&=&\int_{0}^{t}ds\chi _{\alpha \beta }^{\ast }(t,s)L_{\alpha }\Big{\langle}%
\frac{\delta \rho _{st}(t)}{\delta z_{\beta }^{\ast }(s)}\Big{\rangle} 
\notag \\
&&+\int_{0}^{t}ds\eta _{\alpha \beta }(t,s)L_{\alpha }\Big{\langle}\frac{%
\delta \rho _{st}(t)}{\delta z_{\beta }(s)}\Big{\rangle},
\label{Multiplicative1}
\end{eqnarray}%
and similarly%
\begin{eqnarray}
\langle \rho _{st}(t)\mathcal{F}^{\dagger }\left( t\right) \rangle
&=&\int_{0}^{t}ds\chi _{\alpha \beta }(t,s)\Big{\langle}\frac{\delta \rho
_{st}(t)}{\delta z_{\beta }(s)}\Big{\rangle}L_{\alpha }^{\dagger }  \notag \\
&&+\int_{0}^{t}ds\eta _{\alpha \beta }^{\ast }(t,s)\Big{\langle}\frac{\delta
\rho _{st}(t)}{\delta z_{\beta }^{\ast }(s)}\Big{\rangle}L_{\alpha
}^{\dagger }.  \label{Multiplicative2}
\end{eqnarray}%
The convention of sum over repeated subindices applies whenever the
summatory symbol is not written. Given the commutativity of the trace
operation, the indetermination given by Eq. (\ref{trazon}) can be surpassed
by demanding the separability condition\ (\ref{separar}), which due the
analytical property of the wave vector $\left\vert \psi (t)\right\rangle $
implies%
\begin{equation}
\frac{\delta \rho _{st}(t)}{\delta z_{\beta }(s)}=\frac{\delta \left\vert
\psi (t)\right\rangle }{\delta z_{\beta }(s)}\left\langle \psi \left(
t\right) \right\vert ,\;\;\;\;\;\;\;\;\frac{\delta \rho _{st}(t)}{\delta
z_{\beta }^{\ast }(s)}=\left\vert \psi (t)\right\rangle \frac{\delta
\left\langle \psi (t)\right\vert }{\delta z_{\beta }^{\ast }(s)}.
\end{equation}%
Therefore, the contribution $\mathcal{U}[\rho _{st}]$ is determined in an
unique way,%
\begin{equation}
\mathcal{U}[\rho _{st}]\!=\!i\lambda \int\nolimits_{0}^{t}\!ds(L_{\alpha
}^{\dagger }\chi _{\alpha \beta }(t,s)-L_{\alpha }\eta _{\alpha \beta }(t,s))%
\frac{\delta \rho _{st}(t)}{\delta z_{\beta }(s)}\!+\!\mathrm{H.c.},
\label{Usolution}
\end{equation}%
which in turn lead to the general non-Markovian Gaussian stochastic Schr\"{o}%
dinger equation%
\begin{eqnarray}
\frac{d}{dt}\left\vert \psi (t)\right\rangle \! &=&\!-iH_{S}\left\vert \psi
(t)\right\rangle -i\lambda \mathcal{F}\left( t\right) \left\vert \psi
(t)\right\rangle  \label{General} \\
&&\!\!\!-i\lambda \int\nolimits_{0}^{t}\!ds(L_{\alpha }^{\dagger }\chi
_{\alpha \beta }(t,s)-L_{\alpha }\eta _{\alpha \beta }(t,s))\frac{\delta
\left\vert \psi (t)\right\rangle }{\delta z_{\beta }(s)}.  \notag
\end{eqnarray}%
This is the main result of this section. We notice that under the
replacement $L_{\alpha }^{\dagger }\rightarrow L_{\alpha },$ this equation
recovers the stochastic dynamics of Ref. \cite{DiosiPRL}. As demonstrated in
the next section, this replacement is equivalent to a unitary change of the
base of system operators. Hence, Eq. (\ref{General}) and the evolution
obtained in \cite{DiosiPRL} are equivalent dynamics expressed in different
operator bases. On the other hand, when $\eta _{\alpha \beta
}(t,s)\rightarrow 0$ it follows the \textit{standard} non-Markovian quantum
diffusion model \cite{diosi1,budini}. In the next section, we also show that
the underlying symmetries of the generalized time evolution can be inferred
and enlighten from Eq. (\ref{General}) (Sec. III). In addition, this
equivalent evolution allow us to recover in a simple way previous
generalized stochastic dynamics obtained from quantum measurement theory 
\cite{gambetta} (Sec. IV).

\subsection*{Functional structure and density matrix evolution}

As in the standard case \cite{diosi1,budini}, we notice that Eq. (\ref%
{General}) depends on the functional derivative of the wave vector. Most of
the achievements performed in the last years \cite{diosi2,yu,budini,
gamba,lilam,jerarquia,wu,cresser}\ can be applied to the generalized
unraveling. In particular, by following the calculation steps performed in
Ref. \cite{budini} it follows%
\begin{equation}
\frac{\delta \left\vert \psi (t)\right\rangle }{\delta z_{\beta }(s)}%
=-i\lambda G_{st}\left( t,s\right) L_{\beta }\left\vert \psi
(s)\right\rangle ,  \label{functionalSol}
\end{equation}%
where the (functional) propagator is $G_{st}\left( t,s\right) =\left\lceil
\exp -i\int\nolimits_{s}^{t}d\tau \mathcal{T}_{st}\left( \tau \right)
\right\rceil .$ Here, $\left\lceil \cdots \right\rceil $ denotes a time
ordering operation while the functional generator $\mathcal{T}_{st}\left(
t\right) $ defines the time evolution (\ref{General}), that is, $%
(d/dt)\left\vert \psi (t)\right\rangle =\mathcal{T}_{st}\left( t\right)
\left\vert \psi (t)\right\rangle .$ As demonstrated in Ref. \cite{budini}
these expressions allow us to perform consistent approximations (in the
interaction parameter or in the noise correlation times) of the stochastic
wave vector in both a time convoluted and time convolutionless schemes.

By introducing Eq. (\ref{Usolution}) in the stochastic evolution (\ref%
{RhoStochastic}) and after performing the average over realizations [see
Eqs. (\ref{Multiplicative1}) and (\ref{Multiplicative2})], we obtain the
density matrix evolution%
\begin{eqnarray}
\frac{d}{dt}\rho (t) &=&-i[H_{S},\rho (t)]-i\lambda \int_{0}^{t}ds\Big{\{}%
\chi _{\alpha \beta }(t,s)  \label{RhoAverage} \\
&&\times \Big{[}L_{\alpha }^{\dagger },\Big{\langle}\frac{\delta \rho
_{st}(t)}{\delta z_{\beta }(s)}\Big{\rangle}\Big{]}-\mathrm{H.c.}\Big{\}}, 
\notag
\end{eqnarray}%
where the functional derivative is given by%
\begin{equation}
\frac{\delta \rho _{st}(t)}{\delta z_{\beta }(s)}=-i\lambda G_{st}\left(
t,s\right) L_{\beta }\rho _{st}(s)G_{st}^{\dagger }\left( t,s\right) ,
\label{RhoBarSolution}
\end{equation}%
result that follows from Eq. (\ref{functionalSol}).

We notice that Eq. (\ref{RhoAverage}) does not depend explicitly on the
correlations $\eta _{\alpha \beta }(t,s).$ By performing a recursive
expansion in the interaction strength parameter $\lambda $ \cite{budini}, it
is also possible to demonstrate that the density matrix evolution, order by
order in $\lambda ,$ does not depend of $\eta _{\alpha \beta }(t,s).$
Therefore, consistently with Ref. \cite{DiosiPRL}, we conclude that these
correlations only modify the wave vector evolution. Their physical role is
investigated through the following symmetry analysis.

\section{Symmetries}

Here, we analyze the symmetries of the generalized stochastic Schr\"{o}%
dinger evolution defined by Eq. (\ref{General}).

\subsection{Hermitian fluctuations}

When the fluctuation operator (\ref{efe}) is Hermitian%
\begin{equation}
\mathcal{F}^{\dagger }\left( t\right) =\mathcal{F}\left( t\right) ,
\label{efeHermitica}
\end{equation}%
it follows the equality%
\begin{equation}
\sum_{\alpha }L_{\alpha }^{\dagger }\chi _{\alpha \beta }(t,s)=\sum_{\alpha
}L_{\alpha }\eta _{\alpha \beta }(t,s).  \label{iguales}
\end{equation}%
This relation follows by multiplying Eq. (\ref{efeHermitica}) by $z_{\beta
}(s)$ and using the definition of the noise correlations, Eqs. (\ref{corre1}%
) and (\ref{corre2}). We notice that the weight of the functional
contribution in Eq. (\ref{General}) is precisely the difference between the
two terms in the equality (\ref{iguales}). Hence, it consistently vanishes
for Hermitian fluctuations.

\subsection{Invariance under unitary changes of system operator base}

In the derivation of Sec. II the fluctuation operator $\mathcal{F}\left(
t\right) $\ was written in the base defined by the set of operators $%
\{L_{\alpha }\}.$\ Here, we explore which structure assume the evolution (%
\ref{General}) when introducing a new base of operators $\{A_{\mu }\}$
related to the previous one by a unitary transformation. Hence, we write%
\begin{equation}
A_{\mu }=\sum_{\alpha }L_{\alpha }U_{\alpha \mu },\ \ \ \ \ \ \ \ \
L_{\alpha }=\sum_{\mu }A_{\mu }U_{\mu \alpha }^{\dagger },  \label{operators}
\end{equation}%
where the unitary operator $U$\ satisfies $\sum_{\mu }U_{\alpha \mu }U_{\mu
\beta }^{\dagger }=\delta _{\alpha \beta }$ and $\sum_{\alpha }U_{\mu \alpha
}^{\dagger }U_{\alpha \nu }=\delta _{\mu \nu }.$ This change of base allows
to define a new set of noises $\{r_{\mu }(t)\}$ related to the original one
as%
\begin{equation}
r_{\mu }(t)=\sum_{\alpha }U_{\mu \alpha }^{\dagger }z_{\alpha }(t),\ \ \ \ \
\ \ \ z_{\alpha }(t)=\sum_{\mu }U_{\alpha \mu }r_{\mu }(t),  \label{noises}
\end{equation}%
such that the fluctuation operator can be rewritten as%
\begin{equation}
\mathcal{F}\left( t\right) =\sum\limits_{\alpha }z_{\alpha }(t)L_{\alpha
}=\sum_{\mu }r_{\mu }(t)A_{\mu }.  \label{efedosbases}
\end{equation}%
The variational derivative in Eq. (\ref{General}) can be written as%
\begin{equation}
\frac{\delta (\cdot )}{\delta z_{\beta }(s)}\!=\!\sum_{\nu }\int_{0}^{\infty
}d\tau \frac{\delta (\cdot )}{\delta r_{\nu }(\tau )}\frac{\delta r_{\nu
}(\tau )}{\delta z_{\beta }(s)}\!=\!\sum_{\nu }U_{\nu \beta }^{\dagger }%
\frac{\delta (\cdot )}{\delta r_{\nu }(s)},  \label{funcional}
\end{equation}%
where we have used that $\delta r_{\nu }(\tau )/\delta z_{\beta }(s)=\delta
(\tau -s)U_{\nu \beta }^{\dagger }$ [Eq. (\ref{noises})]. Hence, from the
previous expressions, in the new base the general stochastic Schr\"{o}dinger
equation becomes%
\begin{eqnarray}
\frac{d}{dt}\left\vert \psi (t)\right\rangle \! &=&\!-iH_{S}\left\vert \psi
(t)\right\rangle -i\lambda \mathcal{F}\left( t\right) \left\vert \psi
(t)\right\rangle  \label{invariante} \\
&&\!\!\!-i\lambda \int\nolimits_{0}^{t}\!ds(A_{\mu }^{\dagger }\chi _{\mu
\nu }(t,s)-A_{\mu }\eta _{\mu \nu }(t,s))\frac{\delta \left\vert \psi
(t)\right\rangle }{\delta r_{\nu }(s)},  \notag
\end{eqnarray}%
where we have defined the correlations%
\begin{equation}
\chi _{\mu \nu }(t,s)\equiv \sum\limits_{\alpha \beta }U_{\alpha \mu }\chi
_{\alpha \beta }(t,s)U_{\nu \beta }^{\dagger }=\langle r_{\mu }^{\ast
}\left( t\right) r_{\nu }\left( s\right) \rangle ,  \label{NewChi}
\end{equation}%
and also%
\begin{equation}
\eta _{\mu \nu }(t,s)\equiv \sum\limits_{\alpha \beta }U_{\mu \alpha
}^{\dagger }\eta _{\alpha \beta }(t,s)U_{\nu \beta }^{\dagger }=\langle
r_{\mu }\left( t\right) r_{\nu }\left( s\right) \rangle .  \label{NewEta}
\end{equation}%
Equation (\ref{invariante}) and the previous two expressions show that the
general evolution (\ref{General}) is invariant under the unitary changes
defined by Eqs. (\ref{operators}) and (\ref{noises}). One can always choose
an Hermitian base of operators $\{A_{\mu }\}=\{A_{\mu }^{\dagger }\},$ which
explicitly demonstrate that Eq. (\ref{General}) and Eq. (27) in Ref. \cite%
{DiosiPRL}\ are related by a unitary change of operator bases. On the other
hand, from Eq. (\ref{funcional}) it is possible to demonstrate that the
density matrix evolution (\ref{RhoAverage}) is also invariant under the
transformation (\ref{operators}) and (\ref{noises}), that is, it can be
rewritten in terms of the operators $A_{\mu },$ the derivative $\delta \rho
_{st}(t)/\delta r_{\nu }\left( s\right) ,$ and the correlations $\chi _{\mu
\nu }(t,s)$ [Eq. (\ref{NewChi})].

Given that $\chi _{\mu \nu }(t,s)$ and $\eta _{\mu \nu }(t,s)$ only depend
on the correlations $\chi _{\alpha \beta }(t,s)$ and $\eta _{\alpha \beta
}(t,s)$\ respectively, the independence of $\rho (t)$ with respect to $\eta
_{\alpha \beta }(t,s)$ [Eq. (\ref{RhoAverage})] also implies its
independence with respect to $\eta _{\mu \nu }(t,s)$ in the new base.
Therefore, this property is valid in any operator basis, conclusion
consistent with the results of Ref.~\cite{DiosiPRL}.

\subsubsection*{White noises}

For white correlated noises%
\begin{equation}
\chi _{\alpha \beta }(t,s)=\delta (t-s)\delta _{\alpha \beta },\ \ \ \ \ \ \
\ \eta _{\alpha \beta }(t,s)=\delta (t-s)c_{\alpha \beta },  \label{blancos}
\end{equation}%
where $c_{\alpha \beta }$ are complex coefficients, by using Eqs. (\ref%
{RhoAverage}) and (\ref{RhoBarSolution}), the density matrix evolution
becomes a Lindblad equation%
\begin{equation*}
\frac{d\rho (t)}{dt}=-i[H_{S},\rho (t)]+\lambda ^{2}([L_{\alpha },\rho
(t)L_{\alpha }^{\dagger }]+[L_{\alpha }\rho (t),L_{\alpha }^{\dagger }]).
\end{equation*}%
Under the unitary transformation (\ref{operators}), this equation remains
invariant, that is, the only change corresponds to the replacements $%
L_{\alpha }\rightarrow A_{\mu }.$ On the other hand, in addition to this
change, Eq. (\ref{invariante}) results defined by the correlations [see Eqs.
(\ref{NewChi}) and (\ref{NewEta})] 
\begin{subequations}
\begin{eqnarray}
\chi _{\mu \nu }(t,s) &=&\delta (t-s)\delta _{\mu \nu }, \\
\eta _{\mu \nu }(t,s) &=&\delta (t-s)\sum\limits_{\alpha \beta }U_{\mu
\alpha }^{\dagger }c_{\alpha \beta }U_{\nu \beta }^{\dagger }.
\end{eqnarray}%
While the master equation is invariant (does not depend explicitly) under
the unitary transformation $U,$ the stochastic evolution depends on it
through the correlation $\eta _{\mu \nu }(t,s).$ Therefore, if one demand
that both the master equation and the stochastic Schr\"{o}dinger equation
must have the same dependence on $U$ (symmetry) it follows that $c_{\alpha
\beta }\rightarrow 0.$ Hence,$\ \eta _{\alpha \beta }(t,s)\rightarrow 0.$
Based on a different approach, this result was developed in Ref. \cite{rigo}%
. Now, we ask if this kind of arguments are also applicable for the
generalized non-Markovian Schr\"{o}dinger equation.

\subsection{Mapping with microscopic Hamiltonian symmetries}

The previous analysis (valid for white noises) does not apply in the present
framework. In fact, a redefinition of the noises allowed us to conclude that
the generalized Schr\"{o}dinger dynamics (\ref{General}) and the master
equation (\ref{RhoAverage}) share the same invariance symmetry property
under unitary changes of the operator bases. Thus, any constraint applicable
to the noise correlations has to be based on a more fundamental requirement.
With this motivation, we ask if the stochastic Schr\"{o}dinger equation has
the same invariance symmetry property than a microscopic bosonic dynamics
able to induce the same system dynamics.

The total microscopic system-reservoir Hamiltonian $H_{T}$ reads 
\end{subequations}
\begin{equation}
H_{T}=H_{S}+H_{B}+\lambda \sum_{\alpha }L_{\alpha }\otimes Z_{\alpha },
\label{HTotal}
\end{equation}%
where as before $H_{S}$\ is the system Hamiltonian, $H_{B}$ is the bath
Hamiltonian and the remaining contribution gives their interaction.
Similarly, we introduce a unitary change of the system operator base [Eq. (%
\ref{operators})]%
\begin{equation}
A_{\mu }=\sum_{\alpha }L_{\alpha }U_{\alpha \mu },\ \ \ \ \ \ \ \ L_{\alpha
}=\sum_{\mu }A_{\mu }U_{\mu \alpha }^{\dagger },  \label{operatorBIS}
\end{equation}%
and a new set of bath operators%
\begin{equation}
R_{\mu }=\sum_{\alpha }U_{\mu \alpha }^{\dagger }Z_{\alpha },\ \ \ \ \ \ \ \
Z_{\alpha }=\sum_{\mu }U_{\alpha \mu }R_{\mu },  \label{baths}
\end{equation}%
in such a way that the interaction contribution $H_{I}$\ can be written as%
\begin{equation}
H_{I}=\sum_{\alpha }L_{\alpha }\otimes Z_{\alpha }=\sum_{\mu }A_{\mu
}\otimes R_{\mu }.  \label{Hidosbases}
\end{equation}

When tracing out the bath dynamics, $\rho (t)=\mathrm{Tr}_{B}[\exp
(-itH_{T})\rho (0)\otimes \rho _{B}\exp (+itH_{T})],$ where $\mathrm{Tr}%
_{B}[\cdots ]$ is a trace operation in the bath Hilbert space and $\rho _{B}$
is the environment stationary state, given that $\mathrm{Tr}_{B}[\rho
_{B}Z_{\alpha }\left( t\right) ]=\mathrm{Tr}_{B}[\rho _{B}\exp
(+itH_{B})Z_{\alpha }\exp (-itH_{B})]=0,$ the relevant statistical objects
are the bath operator correlations (in an interaction picture with respect
to $H_{B}$) \cite{breuerBook,carmichel}. Therefore, we search under which
conditions both noise correlations $\chi _{\alpha \beta }\left( t,s\right) $
and $\eta _{\alpha \beta }(t,s)$ can be related or mapped with the quantum
bath correlations.

Some conditions are imposed over the bath-noise correlation mapping.

(i) The map has to be invariant under unitary changes of the system operator
base.

(ii) The density matrix evolution obtained from the stochastic and
microscopic Hamiltonian approaches must be the same.

(iii) In addition, the mapping must be consistent, that is, the resulting
noise correlation matrix has to be positive defined [see Eq. (\ref%
{PositiveDefined}) in the Appendix].

Under the previous conditions, different mapping with different physical
motivations can be proposed. A diagonal map arises from a direct mapping
between the stochastic and microscopic dynamics, which lead to the condition 
$\eta _{\alpha \beta }(t,s)\rightarrow 0.$ A non-diagonal map is motivated
from quantum measurement theory, which allows us to raise the previous
condition, $\eta _{\alpha \beta }(t,s)\neq 0,$ if some constraints are
fulfilled. The diagonal mapping can be read as a particular case of the
non-diagonal one. Nevertheless, for clarity, below each case is presented in
a separate way.

\subsubsection{Diagonal correlation mapping}

The stochastic and microscopic density matrix dynamics can be put in
one-to-one correspondence [see Eqs. (\ref{efedosbases}) and (\ref{Hidosbases}%
)] under the following correlation associations 
\begin{subequations}
\label{asociaciones}
\begin{eqnarray}
\chi _{\alpha \beta }\left( t,s\right) &\leftrightarrow &\mathrm{Tr}%
_{B}[\rho _{B}Z_{\alpha }^{\dagger }\left( t\right) Z_{\beta }(s)],
\label{primera} \\
\eta _{\alpha \beta }(t,s) &\leftrightarrow &\mathrm{Tr}_{B}[\rho
_{B}Z_{\alpha }\left( t\right) Z_{\beta }(s)].  \label{segunda}
\end{eqnarray}%
This correlation map is invariant under unitary changes of system operator
base [condition (i)]. In fact, from Eqs. (\ref{noises}) and (\ref{baths}),
in the new base Eq. (\ref{asociaciones}) becomes 
\end{subequations}
\begin{subequations}
\label{RotatedCorrelations}
\begin{eqnarray}
\chi _{\mu \nu }\left( t,s\right) &\leftrightarrow &\mathrm{Tr}_{B}[\rho
_{B}R_{\mu }^{\dagger }\left( t\right) R_{\nu }(s)], \\
\eta _{\mu \nu }(t,s) &\leftrightarrow &\mathrm{Tr}_{B}[\rho _{B}R_{\mu
}\left( t\right) R_{\nu }(s)],
\end{eqnarray}%
which corresponds to the same mapping in the new noise and bath operator
bases.

Given that the average density matrix does not depend on the correlations $%
\eta _{\alpha \beta }(t,s),$ the equality $\chi _{\alpha \beta }\left(
t,s\right) =\mathrm{Tr}_{B}[\rho _{B}Z_{\alpha }^{\dagger }\left( t\right)
Z_{\beta }(s)]$ guarantees condition (ii), that is, the stochastic and
microscopic approach lead to the same density matrix evolution (see Ref. 
\cite{budini}). Notice that due to the Gaussian and bosonic properties of
the fluctuation operator and reservoir respectively, it is not necessary to
define the map for higher correlations.

The map (\ref{asociaciones}) should allows us to associate a microscopic
origin to both noise correlations $\chi _{\alpha \beta }\left( t,s\right) $
and $\eta _{\alpha \beta }(t,s).$ Nevertheless, by taking the relations (\ref%
{asociaciones}) as valid equalities, by using that $H_{I}^{\dag
}(t)=H_{I}(t),$\ from Eq. (\ref{iguales}) one immediately deduce (in an
interaction representation) that the fluctuation operator $\mathcal{F}(t)$
should be Hermitian, implying the cancellation of the variational
contribution in the stochastic Schr\"{o}dinger evolution (\ref{General}).
Hermitian fluctuations can only be mapped with a quantum reservoir at
infinite temperature \cite{budini,rio}. Given an environment at finite
temperature, this contradiction can only be surpassed by imposing the
correlation map 
\end{subequations}
\begin{equation}
\chi _{\alpha \beta }\left( t,s\right) =\mathrm{Tr}_{B}[\rho _{B}Z_{\alpha
}^{\dagger }\left( t\right) Z_{\beta }(s)],  \label{chi}
\end{equation}%
and demanding%
\begin{equation}
\eta _{\alpha \beta }(t,s)=0.  \label{etal}
\end{equation}%
In the Appendix we demonstrate that, without imposing any condition on the
bath properties, these correlations satisfy condition (iii). Therefore, Eq. (%
\ref{chi}) and (\ref{etal}) define a consistent correlation map which in
turn recovers the standard non-Markovian quantum diffusion model \cite%
{diosi1,diosi2} (Eq. (\ref{General}) with $\eta _{\alpha \beta
}(t,s)\rightarrow 0).$ This is one of the main results of this section.
Consistently, Eq. (\ref{etal}) also implies $\eta _{\mu \nu }(t,s)=0.$

The previous correlation map [Eqs. (\ref{chi}) and (\ref{etal})] guaranties
that the stochastic and microscopic evolution (at any temperature) have the
same invariance symmetry property under arbitrary unitary changes of the
system operator base. Notice that the alternative mapping $\eta _{\alpha
\beta }\left( t,s\right) =\mathrm{Tr}_{B}[\rho _{B}Z_{\alpha }\left(
t\right) \,Z_{\beta }(s)]$ with $\chi _{\alpha \beta }\left( t,s\right) =0$
is not consistent with condition (iii) [see Eq. (\ref{PositiveDefined})]. In
fact, for arbitrary complex noises it follows $\chi _{\alpha \beta }\left(
t,s\right) |_{t=s,\alpha =\beta }>0.$

\subsubsection{Non-diagonal (measurement-like) correlation mapping}

In diverse quantum optical arranges, the Markovian dynamics of a system is
inferred from a measurement process performed on the environment. Different
measurement schemes are defined by the quadratures of the bath, which in
turn define different stochastic unravelings \cite{milburnbook,carmichel}
with the same average dynamics. In the present context, a similar degree of
freedom can be introduced by defining a set of \textit{quadrature-like }\cite%
{milburnbook,carmichel} bath operators $\{\mathcal{Z}_{\alpha }(t)\},$ which
read%
\begin{equation}
\mathcal{Z}_{\alpha }(t)\equiv \sum_{\alpha ^{\prime }}M_{\alpha \alpha
^{\prime }}Z_{\alpha ^{\prime }}(t).  \label{GeneralZetal}
\end{equation}%
Here $M_{\alpha \alpha ^{\prime }}$ is an arbitrary complex matrix. When $%
M_{\alpha \alpha ^{\prime }}=\delta _{\alpha \alpha ^{\prime }}$ the
interaction bath operators [see Eq. (\ref{HTotal})] are recovered.
Therefore, consistently with the previous diagonal case, maintaining the
base of system operators $\{L_{\alpha }\},$ instead of Eq. (\ref%
{asociaciones}), we introduce the generalized correlation mapping 
\begin{subequations}
\label{GeneralAssociations}
\begin{eqnarray}
\chi _{\alpha \beta }\left( t,s\right) &=&\mathrm{Tr}_{B}[\rho _{B}\mathcal{Z%
}_{\alpha }^{\dagger }\left( t\right) \mathcal{Z}_{\beta }(s)], \\
\eta _{\alpha \beta }(t,s) &=&\mathrm{Tr}_{B}[\rho _{B}\mathcal{Z}_{\alpha
}\left( t\right) \mathcal{Z}_{\beta }(s)].
\end{eqnarray}

The non-diagonal associations (\ref{GeneralAssociations}) are also invariant
under changes of the system operator base [condition (i)]. In fact, from the
Schr\"{o}dinger evolution (\ref{invariante}), Eq. (\ref{operatorBIS}) and (%
\ref{baths}), it follows Eq. (\ref{RotatedCorrelations}) with the
replacement $R_{\mu }\left( t\right) \rightarrow \mathcal{R}_{\mu }\left(
t\right) ,$ where 
\end{subequations}
\begin{equation}
\mathcal{R}_{\mu }\left( t\right) =M_{\mu \mu ^{\prime }}R_{\mu ^{\prime
}}\left( t\right) =(U_{\mu \alpha }^{\dagger }M_{\alpha \alpha ^{\prime
}}U_{\alpha ^{\prime }\mu ^{\prime }})R_{\mu ^{\prime }}\left( t\right) .
\end{equation}%
Hence, invariance is guaranteed by a unitary transformation of the matrix $%
M_{\alpha \alpha ^{\prime }}.$

Condition (ii) implies that the density matrix evolution has to remains the
same than in the diagonal case, which in fact corresponds to the dynamics
derived from the microscopic Hamiltonian dynamics. Given that in Eqs. (\ref%
{asociaciones}) and (\ref{GeneralAssociations}) the base of system operators
is the same, and given that the density matrix evolution (\ref{RhoAverage})
does not depend on the correlations $\eta _{\alpha \beta }(t)$ \cite%
{DiosiPRL}, it follows the condition%
\begin{equation}
\mathrm{Tr}_{B}[\rho _{B}Z_{\alpha }^{\dagger }\left( t\right) Z_{\beta
}(s)]=\mathrm{Tr}_{B}[\rho _{B}\mathcal{Z}_{\alpha }^{\dagger }\left(
t\right) \mathcal{Z}_{\beta }(s)].  \label{condition}
\end{equation}%
Therefore, the bath operators $\{Z_{\alpha }(t)\}$ and the quadrature-like\
bath operators $\{\mathcal{Z}_{\alpha }\left( t\right) \}$ must have the
same correlations. From Eq. (\ref{GeneralZetal}) this condition explicitly
reads%
\begin{equation*}
\mathrm{Tr}_{B}[\rho _{B}Z_{\alpha }^{\dagger }\left( t\right) Z_{\beta
}(s)]=M_{\alpha ^{\prime }\alpha }^{\dagger }M_{\beta \beta ^{\prime }}%
\mathrm{Tr}_{B}[\rho _{B}Z_{\alpha ^{\prime }}^{\dagger }\left( t\right)
Z_{\beta ^{\prime }}(s)].
\end{equation*}

Even when the previous constraint is fulfilled, it is not possible to
guaranty that condition (iii) is satisfied. That is, it is not possible to
know in general if the quantum correlations defined by $\{\mathcal{Z}%
_{\alpha }\left( t\right) \}$\ [Eq. (\ref{GeneralAssociations})] satisfy or
not the positivity condition given by Eq. (\ref{PositiveDefined}).

In general, it may be difficult to satisfy the previous requirements without
imposing some condition on the matrix $M_{\alpha \beta }$ or alternatively
on the properties of the bath. Notice that in the white noise approximation $%
\mathrm{Tr}_{B}[\rho _{B}Z_{\alpha }^{\dagger }\left( t\right) Z_{\beta
}(s)]=\delta (t-s)\delta _{\alpha \beta }$ [Eq. (\ref{blancos})], Eq. (\ref%
{condition}) is automatically satisfied for unitary matrixes, $M_{\alpha
^{\prime }\alpha }^{\dagger }M_{\beta \alpha ^{\prime }}=\delta _{\alpha
\beta }$ (Stratonovich calculus).

Assuming the consistence of the map (\ref{GeneralAssociations}) with
conditions (ii) and (iii), in contrast to the diagonal case, here the
Hermiticity of the interaction Hamiltonian $H_{I}(t)$ and Eq. (\ref{iguales}%
) do not lead in general to any contradiction. Therefore, given that Eqs. (%
\ref{GeneralAssociations}) are consistent with the correlations of a set of
complex noises, and given that Eq. (\ref{condition}) is satisfied, the non
diagonal mapping [Eq. (\ref{GeneralZetal})] allow us to define the
correlations of the generalized Eq. (\ref{General}) from the microscopic
underlying evolution. This is the second main result of this section. In
fact, we obtained conditions under which the different unravelings related
to Eq. (\ref{General}) can be associated to the microscopic Hamiltonian
dynamics. At this stage, notice that the diagonal case can be read as a
particular case of the present one.

\section{Examples}

In the next examples we analyze the consequences of the previous constraints
when applied to different system-bath interaction structures. The relation
of the generalized stochastic unraveling with previous results based on
quantum-measurement theory \cite{gambetta} is also revisited.

\subsection{Hermitian single channel}

The general evolution (\ref{General}) takes a simpler form when one take
into account only one single channel, $\mathcal{F}\left( t\right) =z(t)L,$
defined by an Hermitian operator, $L^{\dag }=L.$ Therefore, it follows%
\begin{eqnarray}
\frac{d}{dt}\left\vert \psi (t)\right\rangle \! &=&\!-iH_{S}\left\vert \psi
(t)\right\rangle -i\lambda z(t)L\left\vert \psi (t)\right\rangle
\label{OneChanel} \\
&&\!\!\!-i\lambda L\int\nolimits_{0}^{t}\!ds(\chi (t,s)-\eta (t,s))\frac{%
\delta \left\vert \psi (t)\right\rangle }{\delta z(s)}.  \notag
\end{eqnarray}%
From a phenomenological point of view, one can propose different noise
correlations $\chi (t,s)$ and $\eta (t,s),$ as for example exponential ones.
In this case, it is possible to explicitly show [see Appendix, Eq. (\ref%
{Diferencia})] that the extra correlation $\eta (t,s)$ allows to smoothly
change the stochastic unraveling between the standard non-Markovian quantum
diffusion model $[\eta (t,s)=0]$ and a pure stochastic Hamiltonian $[\eta
(t,s)=\chi (t,s)].$ On the other hand, the consistency of this evolution
with the invariance symmetry of a microscopic Hamiltonian description is
analyzed below.

Eq. (\ref{OneChanel}) should be associated with an interaction Hamiltonian
of the form $H_{I}=L\otimes Z,$ where the bath operator is Hermitian, $%
Z=Z^{\dag },$ with (complex) correlation $\mathrm{Tr}_{B}[\rho _{B}Z\left(
t\right) Z(s)].$ If one try to impose the associations (\ref{asociaciones}),
it follows that $\chi (t,s)=\eta (t,s).$ This equality can only be satisfied
by a real noise $z(t)$ with a real correlation [see Eq. (\ref%
{PositiveDefined})]. Consequently $\mathcal{F}^{\dagger }\left( t\right) =%
\mathcal{F}\left( t\right) .$ Given that $\chi (t,s)=\chi ^{\ast }(t,s)=\eta
(t,s)=\eta ^{\ast }(t,s),$ it also implies $\mathrm{Tr}_{B}[\rho _{B}Z\left(
t\right) Z(s)]=\mathrm{Tr}_{B}[\rho _{B}Z\left( s\right) Z(t)].$ In general,
this equality can only be satisfied when $\rho _{B}$ is proportional to the
identity matrix, that is, a thermal state at infinite temperature. This
contradiction, which also implies the cancellation of the functional
contribution in Eq. (\ref{OneChanel}), is raised up by taking $\chi (t,s)=%
\mathrm{Tr}_{B}[\rho _{B}Z\left( t\right) Z(s)]$ and $\eta (t,s)=0,$ that
is, consitently we recover the diagonal correlation mapping defined by Eqs. (%
\ref{chi}) and (\ref{etal}) respectively.

Given that the interaction Hamiltonian only involves one single Hermitian
operator $Z,$ here it is not possible to introduce a non-diagonal
correlation mapping (\ref{GeneralAssociations}). Therefore, the correlation $%
\eta (t,s)$ in Eq.~(\ref{OneChanel}) \textit{cannot} be related to the
microscopic interaction if the same invariance symmetry property is demanded.

\subsubsection*{Dephasing dynamics}

Dephasing dynamics is one particular physical case of the previous
situation. It emerges when the (Hermitian) coupling operator commutates with
the system Hamiltonian, $[H_{S},L]=0.$ Hence, in Eq. (\ref{OneChanel}), the
variational derivative can be written as%
\begin{equation}
\frac{\delta \left\vert \psi (t)\right\rangle }{\delta z(s)}=-i\lambda
L\left\vert \psi (t)\right\rangle ,  \label{FunctionalDephasing}
\end{equation}%
result that follows straightforward from Eq. (\ref{functionalSol}). On the
other hand, from Eqs. (\ref{RhoAverage}) and (\ref{RhoBarSolution}) it
follows the density matrix evolution%
\begin{equation}
\frac{d}{dt}\rho (t)=-i[H_{S},\rho (t)]-\lambda
^{2}\int\nolimits_{0}^{t}ds\kappa (t,s)[L,[L,\rho (t)]],
\end{equation}%
where $\kappa (t,s)=[\chi (t,s)+\chi ^{\ast }(t,s)]/2.$ We notice that this
time evolution also arises from a pure stochastic Hamiltonian (see Eq. (40)
in Ref. \cite{rio}) defined with a real Gaussian noise with correlation $%
\kappa (t,s).$ Therefore, all the set of ensembles defined by Eq. (\ref%
{OneChanel}) and (\ref{FunctionalDephasing}) are equivalent in average to a
stochastic Hamiltonian. This property is stretchy related to the dephasing
property of the underlying fluctuations.

\subsection{Quantum optical-like microscopic interaction}

For quantum optical systems, where a Markovian approximation applies, the
correlations $\eta _{\alpha \beta }(t,s)$ (delta correlated) define
different ensemble dynamics associated to different measurement processes
performed over the environmental degrees of freedom \cite{howard}. Based on
quantum measurement theory, Gambetta and Wiseman demonstrated that a similar
result is valid in the non-Markovian regime \cite{gambetta}. Below, we show
that those results can be recovered as diagonal and non-diagonal correlation
mappings associated to the same microscopic Hamiltonian. In fact, the degree
of freedom associated to the different measurement processes here can be
related to alternative definitions of the bath operators $\{\mathcal{Z}%
_{\alpha }(t)\}$ [Eq. (\ref{GeneralZetal})].

The system Hamiltonian is split as $H_{S}=H_{0}+H.$ The bath Hamiltonian is
taken as $H_{B}=\sum_{k}\omega _{k}a_{k}^{\dagger }a_{k},$ where $%
a_{k}^{\dagger }$ and $a_{k}$ are bosonic creation and annihilation
operators respectively, $\omega _{k}$ the frequency of each mode. The
system-bath interaction is $H_{I}=i(L\otimes b^{\dagger }-L^{\dagger
}\otimes b),$ with $b=\sum_{k}g_{k}a_{k},$ where $g_{k}$ are coupling
constants. The stationary bath density matrix $\rho _{B}$ is the vacuum
state. In an interaction representation with respect to $H_{0}$ and $H_{B},$
it follows%
\begin{equation}
H_{I}(t)=(iL)Z^{\dagger }(t)+(-iL^{\dagger })Z(t),
\label{InteractionHamiltonian}
\end{equation}%
where the bath operator is $Z(t)=\sum_{k}g_{k}a_{k}e^{-i\Omega _{k}t},$ with 
$\Omega _{k}\equiv \omega _{k}-\omega _{0}.$ The frequency $\omega _{0}$
follows from the assumption $L(t)=e^{+itH_{0}}Le^{-itH_{0}}=Le^{-i\omega
_{0}t}.$

Noticing that $\{Z_{\alpha }(t)\}=\{Z^{\dagger }(t),Z(t)\}$ characterize the
interaction Hamiltonian in the base of system operators $\{L_{\alpha
}\}=\{iL,-iL^{\dagger }\},$ demanding the positivity of the noise
correlation matrix [see Eq. (\ref{PositiveDefined})], it is simple to check
that the \textit{diagonal} correlation mapping defined by Eqs. (\ref{chi})
and (\ref{etal}) can only be consistently satisfied by introducing one
single complex noise $z_{c}(s),$ $\mathcal{F}\left( t\right) =z_{c}(t)(iL),$
such that $\chi _{c}(t,s)=\left\langle z_{c}^{\ast }(t)z_{c}(s)\right\rangle
=\mathrm{Tr}_{B}[\rho _{B}Z(t)Z^{\dagger }\left( s\right)
]=\sum_{k}|g_{k}|^{2}e^{-i\Omega _{k}(t-s)},$ and $\eta
_{c}(t,s)=\left\langle z_{c}(t)z_{c}(s)\right\rangle =0.$ Hence, Eq. (\ref%
{General}) becomes $[H(t)=e^{+itH_{0}}He^{-itH_{0}}]$%
\begin{eqnarray}
\frac{d}{dt}\left\vert \psi (t)\right\rangle \! &=&\!-iH(t)\left\vert \psi
(t)\right\rangle +z_{c}(t)L\left\vert \psi (t)\right\rangle
\label{coherente} \\
&&\!\!\!-L^{\dagger }\int\nolimits_{0}^{t}\!ds\chi _{c}(t-s)\frac{\delta
\left\vert \psi (t)\right\rangle }{\delta z_{c}(s)}.  \notag
\end{eqnarray}%
From quantum measurement theory, this evolution can be read as a \textit{%
coherent unravelling} of the bath (see Eq. (3.22) in \cite{gambetta}) where $%
Z(t)$ is the noise operator. A heterodyne measurement process is recovered
in the Markovian limit \cite{gambetta}.

As an example of \textit{non-diagonal} mapping, Eq. (\ref{GeneralZetal}), we
take the single operator $\mathcal{Z}(t)=Z(t)+Z^{\dagger }(t),$ and
consistently with the diagonal case $\mathcal{F}\left( t\right)
=z_{q}(t)(iL).$ It is simple to check that Eq. (\ref{condition}) is
fulfilled. In fact, given that $\mathcal{Z}^{\dagger }(t)=\mathcal{Z}(t),$
Eq. (\ref{GeneralAssociations}) leads to the correlations $\chi
_{q}(t,s)=\eta _{q}(t,s)=\mathrm{Tr}_{B}[\rho _{B}\mathcal{Z}(t)\mathcal{Z}%
(s)]=\sum_{k}|g_{k}|^{2}e^{-i\Omega _{k}(t-s)}$ (the same correlation than
in the coherent unraveling). Nevertheless, these correlations do not have
associated a positive covariance matrix [Eq. (\ref{PositiveDefined})]. This
contradiction can be raised up by taking the complex noise $z_{q}(t)$ as a
real one, implying real correlations. Given that $L\neq L^{\dagger },$ here
a real noise does not implies $\mathcal{F}^{\dagger }(t)=\mathcal{F}(t).$
The previous complex correlations become real if we demand extra properties
to the bath. If $\Omega _{-k}=-\Omega _{k}$ and, given that $H_{I}^{\dagger
}(t)=H_{I}(t),$ demanding $g_{-k}=g_{k}^{\ast },$ it follows $\chi
_{q}(t,s)=\eta _{q}(t,s)=2\sum_{k>0}|g_{k}|^{2}\cos [\Omega _{k}(t-s)]\equiv
\beta _{q}(t-s).$ Therefore, from Eq. (\ref{General}) we arrive to the
generalized Schr\"{o}dinger equation%
\begin{eqnarray}
\frac{d}{dt}\left\vert \psi (t)\right\rangle \! &=&\!-iH(t)\left\vert \psi
(t)\right\rangle +z_{q}(t)L\left\vert \psi (t)\right\rangle
\label{quadratura} \\
&&\!\!\!-(L^{\dagger }+L)\int\nolimits_{0}^{t}\!ds\beta _{q}(t-s)\frac{%
\delta \left\vert \psi (t)\right\rangle }{\delta z_{q}(s)}.  \notag
\end{eqnarray}%
This evolution can be read as a \textit{quadrature unravelling} of the bath
(see Eq. (4.30) in \cite{gambetta}) where $Z(t)+Z^{\dagger }(t)$ is the
noise operator. A homodyne measurement process is recovered in the Markovian
limit \cite{gambetta}.

We derived Eqs. (\ref{coherente}) and (\ref{quadratura}) on the basis of the
diagonal and non-diagonal correlation mapping introduced previously. These
results give a solid support to the present and previous analysis \cite%
{gambetta,hidden}, which rely on quantum measurement theory.

Based on the analysis of Refs. \cite{gambetta,hidden}, we also conclude that
the ensemble of realizations associated to generalized Schr\"{o}dinger
dynamics such as Eq. (\ref{General}) cannot be read as \textit{conditional}
states of a system subjected to a \textit{continuous} measurement process
over the environment degrees of freedom. In fact, as argued in Refs. \cite%
{gambetta,hidden}, in the non-Markovian regime the wave vector only
represents the state the system would be at a single time. The property of
linking solutions at different times to make a trajectory of a continuously
monitored system is lost \cite{Diosol}.

\section{Summary and conclusions}

We based the present analysis on a stochastic approach. Postulating an
underlying evolution with multiplicative noises, we derived the generalized
non-Markovian Gaussian stochastic Schr\"{o}dinger equation (\ref{General})
from the condition of average trace preservation. Complex Gaussian noises
with arbitrary correlations $\chi _{\alpha \beta }(t,s)$ and $\eta _{\alpha
\beta }(t,s),$ Eqs. (\ref{corre1}) and (\ref{corre2}) respectively, were
considered.

We focused our study on the symmetries of the obtained evolution. The
generalized Schr\"{o}dinger equation, after a redefinition of the noises, is
invariant under arbitrary unitary changes of the system operator base. This
property allowed us to conclude that Eq. (\ref{General}) and the stochastic
dynamics derived in Ref. \cite{DiosiPRL} are related by a unitary
transformation. On the other hand, the associated master equation share the
same symmetry property than the stochastic Schr\"{o}dinger equation.
Therefore, in contrast to previous analysis with Markovian dynamics, we
asked under which conditions the invariance property can be mapped with that
of a microscopic (bosonian) system-bath dynamics, which in turn lead us to
consider a mapping between the noise correlations $\chi _{\alpha \beta
}(t,s) $ and $\eta _{\alpha \beta }(t,s)$ with the bath operator
correlations.

Two kind of maps were introduced. In the diagonal one, the quantum reservoir
correlations are defined by the bath operators corresponding to the
system-environment interaction Hamiltonian. In the non-diagonal map, a new
set of bath operators is introduced (quadrature-like bath operators), being
related to the previous ones by an arbitrary linear transformation.

For the diagonal correlation mapping [Eq. (\ref{asociaciones})], the
invariance symmetry property of the microscopic dynamics is shared by the
stochastic unravelling only when the correlation $\eta _{\alpha \beta }(t,s)$%
\ vanishes, Eqs. (\ref{chi}) and (\ref{etal}), recovering in consequence the
standard non-Markovian quantum diffusion model \cite{diosi1}. For the
non-diagonal correlation map [Eq. (\ref{GeneralAssociations})], the
invariance symmetry is shared only if some constraints are fulfilled. In
fact, the independence of the density matrix evolution with respect to $\eta
_{\alpha \beta }(t,s)$ and the positivity of the noise correlation matrix
lead to the necessary conditions Eqs. (\ref{condition}) and (\ref%
{PositiveDefined}) respectively. These requirements establish the conditions
under which the noise correlations that set the generalized unraveling, $%
\chi _{\alpha \beta }(t,s)$ and $\eta _{\alpha \beta }(t,s),$ can be defined
from the properties of the microscopic system-bath interaction.

As example, we considered the case of a single noise channel defined by an
Hermitian operator [Eq. (\ref{OneChanel})]. We concluded that only the
standard version of the dynamics can be related with a microscopic
description if the invariance property is demanded. On the other hand,
previous dynamics derived from quantum measurement theory were recovered
from diagonal and non-diagonal correlation mapping [Eqs. (\ref{coherente})
and (\ref{quadratura}) respectively] applied to the same optical-like
system-environment interaction \cite{gambetta}. These examples show that the
consistence of the non-diagonal correlation map put severe constraints on
the bath properties, which in general may or not be fulfilled.

The present analysis not only clarifies the recent advances in the
formulation of Gaussian stochastic wave vector dynamics \cite{DiosiPRL} but
also define the constraints under which the generalized unraveling can be
put in one-to-one correspondence with a microscopic unitary description.
They also supports the stringent constraints on the interpretation of
non-Markovian Schr\"{o}dinger equations as conditional measurement states at
different times \cite{Diosol}.

\section*{Acknowledgment}

This work was supported by Consejo Nacional de Investigaciones Cient\'{\i}%
ficas y T\'{e}cnicas (CONICET), Argentina.

\appendix*

\section{Arbitrary complex Gaussian noises}

The statistical properties of an arbitrary set of complex Gaussian noises $%
\{z_{\alpha }(t)\}$ can be defined through the characteristic functional%
\begin{equation}
G[f,g]\equiv \left\langle \exp \left\{ i\int_{0}^{t}d\tau \lbrack f_{\alpha
}(\tau )z_{\alpha }(\tau )+g_{\alpha }(\tau )z_{\alpha }^{\ast }(\tau
)]\right\} \right\rangle ,\ \ \ \   \label{Ge}
\end{equation}%
where $f$ and $g$ denotes a set of test functions $\{f_{\alpha }(t)\}$ and $%
\{g_{\alpha }(t)\}.$ The Gaussian statistics implies%
\begin{eqnarray}
\ln G[f,g] &=&-\int_{0}^{t}d\tau \int_{0}^{\tau }dsg_{\alpha }(\tau )\chi
_{\alpha \beta }(\tau ,s)f_{\beta }(s)  \notag \\
&&-\int_{0}^{t}d\tau \int_{0}^{\tau }dsf_{\alpha }(\tau )\chi _{\alpha \beta
}^{\ast }(\tau ,s)g_{\beta }(s)  \notag \\
&&-\int_{0}^{t}d\tau \int_{0}^{\tau }dsf_{\alpha }(\tau )\eta _{\alpha \beta
}(\tau ,s)f_{\beta }(s)  \notag \\
&&-\int_{0}^{t}d\tau \int_{0}^{\tau }dsg_{\alpha }(\tau )\eta _{\alpha \beta
}^{\ast }(\tau ,s)g_{\beta }(s).\ \ \ \ \ \   \label{lnG}
\end{eqnarray}%
By functional derivation it follows that $\langle z_{\alpha }\left( t\right)
\rangle =0,$ and the correlations defined by Eqs. (\ref{corre1}) and (\ref%
{corre2}). These definitions cover the particular cases of real noises as
well as the case where $\eta _{\alpha \beta }(t,s)=0.$

\textit{Novikov's theorem} \cite{novikov} gives an exact result for the mean
value of the product between a Gaussian noise and any functional $\mathcal{M}
$ of it. This theorem can be generalized for the set of complex noises
defined by Eq. (\ref{lnG}). Using the Gaussian property and the correlation
definition it is possible to obtain%
\begin{eqnarray}
\langle z_{\gamma }(t)\mathcal{M}[\{z_{\alpha }(t)\}]\rangle \!\!
&=&\!\!\int_{0}^{t}ds\chi _{\gamma \beta }^{\ast }(t,s)\Big{\langle}\frac{%
\delta \mathcal{M}[\{z_{\alpha }(t)\}]}{\delta z_{\beta }^{\ast }(s)}%
\Big{\rangle}  \label{Novikov} \\
&&+\int_{0}^{t}ds\eta _{\gamma \beta }(t,s)\Big{\langle}\frac{\delta 
\mathcal{M}[\{z_{\alpha }(t)\}]}{\delta z_{\beta }(s)}\Big{\rangle},\ \ \  
\notag
\end{eqnarray}%
where $\mathcal{M}[\{z_{\alpha }(t)\}]$ denotes a functional that depends on
the set of noises $\{z_{\alpha }(t)\}.$ The noise $z_{\gamma }(t)$ belongs
to this set. This generalized Novikov theorem has been applied in the
derivation of Sec. II.

\subsection*{Correlation matrix}

The correlation matrix that define Eq. (\ref{lnG}) has to satisfy a
positivity constraint. By defining the scalar $w\equiv
\sum_{i=1}^{n}(a_{i}^{\alpha }z_{\alpha }(t_{i})+b_{i}^{\alpha }z_{\alpha
}^{\ast }(t_{i})),$ where $a_{i}^{\alpha }$ and $b_{i}^{\alpha }$\ are
arbitrary coefficients, $n\in \mathbf{N},$ $\{t_{i}\}$ arbitrary times, the
absolute value $\left\vert w\right\vert ^{2}\geq 0$ implies that the four
blocks correlation kernel 
\begin{subequations}
\label{PositiveDefined}
\begin{equation}
\left( 
\begin{array}{cc}
\mathbf{\chi } & \mathbf{\eta }^{\ast } \\ 
\mathbf{\eta } & \mathbf{\chi }^{\ast }%
\end{array}%
\right) \geq 0,
\end{equation}%
must be positive defined, where $\mathbf{\chi }\leftrightarrow \chi _{\alpha
\beta }(t_{i},t_{j})$ and $\mathbf{\eta }\leftrightarrow \eta _{\alpha \beta
}(t_{i},t_{j}).$ An equivalent condition follows with $w\equiv
\sum_{i=1}^{n}(a_{i}^{\alpha }\mathrm{Re}[z_{\alpha }(t_{i})]+b_{i}^{\alpha }%
\mathrm{Im}[z_{\alpha }(t_{i})]),$ leading to%
\begin{equation}
\left( 
\begin{array}{cc}
\mathrm{Re}[\mathbf{\chi }+\mathbf{\eta }] & \mathrm{Im}[\mathbf{\chi }+%
\mathbf{\eta }] \\ 
-\mathrm{Im}[\mathbf{\chi }-\mathbf{\eta }] & \mathrm{Re}[\mathbf{\chi }-%
\mathbf{\eta }]%
\end{array}%
\right) \geq 0.
\end{equation}%
In both cases, when $n=1$ the positivity of the variance matrix of $%
\{z_{\alpha }(t)\}$ is recovered. Notice that $\eta _{\alpha \beta
}(t,s)=\chi _{\alpha \beta }(t,s)$ is an admissible correlation matrix only
for real noises.

\subsubsection{Bosonic bath correlation matrix}

The correlation matrixes $\chi _{\alpha \beta }\left( t,s\right) $ and $\eta
_{\alpha \beta }\left( t,s\right) $ corresponding to the diagonal map, Eq. (%
\ref{chi}) and (\ref{etal}) respectively, satisfy the positivity constraint (%
\ref{PositiveDefined}). This property is fulfilled without imposing any
special constraint on the bath properties or on the interaction Hamiltonian.
In fact, by writing the interaction bath operators as linear combinations of
free bosonic modes 
\end{subequations}
\begin{equation}
Z_{\alpha }(t)=\int d\omega (g_{\alpha \alpha ^{\prime }}(\omega )b_{\alpha
^{\prime }\omega }^{\dagger }e^{+i\omega t}+h_{\alpha \alpha ^{\prime
}}(\omega )b_{\alpha ^{\prime }\omega }e^{-i\omega t}),  \label{Zetoles}
\end{equation}%
with commutation relations $[b_{\alpha \omega },b_{\beta \omega ^{\prime
}}^{\dagger }]=\delta _{\alpha \beta }\delta (\omega -\omega ^{\prime }),$
it follows%
\begin{eqnarray}
\chi _{\alpha \beta }\left( t,s\right) &=&\int d\omega (n_{\gamma \omega
}+1)g_{\alpha \gamma }^{\ast }(\omega )g_{\beta \gamma }(\omega )e^{-i\omega
(t-s)}  \notag \\
&&+\int d\omega n_{\gamma \omega }h_{\alpha \gamma }^{\ast }(\omega
)h_{\beta \gamma }(\omega )e^{+i\omega (t-s)},\ \ \ \ \ 
\end{eqnarray}%
where $n_{\gamma \omega }$ is the average thermal number in each mode. Given
that $\eta _{\alpha \beta }\left( t,s\right) =0,$ the positivity constraint (%
\ref{PositiveDefined}) (at any bath temperature) is satisfied whenever the
matrixes of complex coefficients $[(n_{\gamma \omega }+1)g_{\alpha \gamma
}^{\ast }(\omega )g_{\beta \gamma }(\omega )]$ and $[n_{\gamma \omega
}h_{\alpha \gamma }^{\ast }(\omega )h_{\beta \gamma }(\omega )]$ are
positive defined, that is, the bath spectrum matrix is positive defined. On
the other hand, we notice that non-stationary correlations \cite{DiosiPRL} $%
\chi _{\alpha \beta }\left( t,s\right) \neq \chi _{\alpha \beta }\left(
t-s\right) ,$ only arise if the underlying Hamiltonian (\ref{HTotal}) is
time dependent. For example, the interaction strength may be time dependent, 
$\lambda \rightarrow \lambda \varphi (t),$ $\chi _{\alpha \beta }\left(
t,s\right) \rightarrow \varphi ^{2}(t)\chi _{\alpha \beta }\left( t-s\right)
,$ or alternatively the previous interaction operators (\ref{Zetoles}) are
defined with time dependent coefficients $g_{\alpha \alpha ^{\prime
}}(\omega )\rightarrow g_{\alpha \alpha ^{\prime }}(\omega ,t),$ $h_{\alpha
\alpha ^{\prime }}(\omega )\rightarrow h_{\alpha \alpha ^{\prime }}(\omega
,t).$

\subsubsection{Exponential noise correlations}

As an (one dimensional) example, we assume that the realizations of the
noise $z(t)$ obey the linear stochastic differential equation%
\begin{equation}
\frac{d}{dt}z(t)=-(\gamma +i\Omega )z(t)+\xi (t),  \label{ele}
\end{equation}%
where $\gamma >0$ and $\Omega $ are real parameters. The complex Gaussian
white noise $\xi (t)$ satisfies 
\begin{subequations}
\begin{eqnarray}
\left\langle \xi ^{\ast }(t)\xi (s)\right\rangle &=&D\delta (t-s), \\
\left\langle \xi (t)\xi (s)\right\rangle &=&D^{\prime }\delta (t-s),
\end{eqnarray}%
where $D^{\prime }\leq D$ are also real parameters. By integrating Eq. (\ref%
{ele}) as $z(t)=z(0)\exp [-(\gamma +i\Omega )t]+\int_{0}^{t}dt^{\prime }\exp
[-(\gamma +i\Omega )(t-t^{\prime })]\xi (t^{\prime }),$ and assuming
stationary initial conditions for $z(t),$ it follows the \textit{exponential
correlations} 
\end{subequations}
\begin{eqnarray*}
\chi (t,s) &=&\left\langle z^{\ast }(t)z(s)\right\rangle =\frac{D}{2\gamma }%
\exp [-(\gamma -i\Omega )(t-s)], \\
\eta (t,s) &=&\left\langle z(t)z(s)\right\rangle =\frac{D^{\prime }}{%
2(\gamma +i\Omega )}\exp [-(\gamma +i\Omega )(t-s)],
\end{eqnarray*}%
where $t\geq s.$ In a Fourier domain, it is possible to demonstrate that
these objects obey Eq. (\ref{PositiveDefined}).

The degree of freedom introduced by $\eta (t,s)$ in Eq. (\ref{OneChanel})
can be easily read by taking $\Omega =0,$ leading to%
\begin{equation}
\chi (t,s)-\eta (t,s)=\Big{(}1-\frac{D^{\prime }}{D}\Big{)}\frac{D}{2\gamma }%
\exp [-\gamma (t-s)].  \label{Diferencia}
\end{equation}%
Therefore, the dimensionless parameter $D^{\prime }/D$ allows to
continuously departs from a Hamiltonian stochastic dynamics $(D^{\prime }=D)$
and, in the other extreme, to reach the standard non-Markovian diffusion
model, that is, $D^{\prime }=0.$ While this property was derived for this
particular case, it is simple to realize that the inclusion of the extra
correlations $\eta _{\alpha \beta }(t,s)$ allow to smoothly reach these
limits when considering the general evolution (\ref{General}) with an
Hermitian operator base, $\{L_{\alpha }^{\dagger }\}=\{L_{\alpha }\}.$


\begin{thebibliography}{99}
\bibitem{breuerBook} H. P. Breuer and F. Petruccione, \textit{Theory of Open
Quantum Systems} (Oxford University Press, Oxford, England, 2002).

\bibitem{barchiellibook} A. Barchielli and M. Gregoratti, \textit{Quantum
Trajectories and Measurements in Continuous time---The diffusive case},
Lectures Notes in Physics Vol. \textbf{782} (Springer, Berlin, 2009).

\bibitem{milburnbook} H. Wiseman and G. Milburn, \textit{Quantum Measurement
and Control} (Cambridge University Press, Cambridge, England, 2010).

\bibitem{carmichel} H. J. Carmichel, \textit{An Open Systems approach to
Quantum Optics} (Lectures Notes in Physics, Springer-Verlag, 1993).

\bibitem{plenio} M. B. Plenio and P. L. Knight, \textit{The quantum jump
approach to dissipative dynamics in quantum optics}, Rev. Mod. Phys. \textbf{%
70}, 101 (1998).

\bibitem{gisin} N. Gisin and I. C. Percival, The quantum-state diffusion
model applied to open systems, J. Phys. A: Math. Gen. \textbf{25}, 5677
(1992); M. Rigo and N. Gisin, Unravellings of the master equation and the
emergence of a classical world, Quantum Semiclass. Opt. \textbf{8}, 255
(1996); N. Gisin and M. B. Cibils, Quantum diffusions, quantum dissipation
and spin relaxation, J. Phys. A \textbf{25}, 5165 (1992); L. Di\'{o}si,
Quantum stochastic processes as models for state vector reduction, J. Phys.
A \textbf{21}, 2885 (1988).

\bibitem{rigo} M. Rigo, F. Mota-Furtado, and P. F. O'Mahony, Continuous
stochastic Schr\"{o}dinger equations and localization, J. Phys. A: Math.
Gen. \textbf{30}, 7557 (1997).

\bibitem{howard} H. M. Wiseman and L. Di\'{o}si, Complete parametrization,
and invariance, of diffusive quantum trajectories for Markovian open
systems, Chem. Phys. \textbf{268}, 91 (2001).

\bibitem{ghirardi} A. Bassi and G. C. Ghirardi, \textit{Dynamical reduction
models}, Phys. Rep. \textbf{379}, 257 (2003); N. Gisin and I. C. Percival,
Quantum state diffusion, localization and quantum dispersion entropy, J.
Phys. A: Math. Gen. \textbf{26}, 2233 (1993); N. Gisin and I. C. Percival,
The quantum state diffusion picture of physical process, J. Phys. A: Math.
Gen. \textbf{26}, 2245 (1993).

\bibitem{diosi1} L. Di\'{o}si and W. T. Strunz, The non-Markovian stochastic
Schr\"{o}dinger equation for open systems, Phys. Lett. A \textbf{235}, 569
(1997); W. T. Strunz, Linear quantum state diffusion for non-Markovian open
quantum systems, Phys. Lett. \textbf{224}, 25 (1996); L. Di\'{o}si, Exact
semiclassical wave equation for stochastic quantum optics, Quantum
Semiclass. Opt. \textbf{8}, 309 (1996).

\bibitem{diosi2} L. Di\'{o}si, N. Gisin, and W. T. Strunz, Non-Markovian
quantum state diffusion, Phys. Rev. A \textbf{58}, 1699 (1998).

\bibitem{yu} T. Yu, L. Di\'{o}si, N. Gisin and W. T. Strunz, Non-Markovian
quantum-state diffusion: Perturbation approach, Phys. Rev. A \textbf{60}, 91
(1999).

\bibitem{cresser} J. D. Cresser, A Heisenberg Equation-of-motion Derivation
of Stochastic Schr\"{o}dinger Equations for Non-Markovian Open Systems,
Laser Phys. \textbf{10}, 1 (2000).

\bibitem{budini} A. A. Budini, Non-Markovian Gaussian stochastic wave
vector, Phys. Rev. A \textbf{63}, 012106 (2000).

\bibitem{gamba} J. Gambetta and H. M. Wiseman, Perturbative approach to
non-Markovian stochastic Schr\"{o}dinger equations, Phys. Rev. A \textbf{66}%
, 052105 (2002).

\bibitem{lilam} Z. Li, C. Yip, H. Deng, M. Chen, T. Yu, J. Q. You, and C.
Lam, Approach to solving spin-boson dynamics via non-Markovian quantum
trajectories, Phys. Rev. A \textbf{90}, 022122 (2014).

\bibitem{jerarquia} D. Suess, A. Eisfeld, and W. T. Strunz, Hierarchy of
Stochastic Pure States for Open Quantum System Dynamics, Phys. Rev. Lett. 
\textbf{113}, 150403 (2014).

\bibitem{wu} J. Xu, X. Zhao, J. Jing, L. Wu, and T. Yu, Perturbation methods
for non-Markovian quantum state diffusion equation, J. Phys. A \textbf{47},
435301 (2014).

\bibitem{strunz} W. T. Strunz, L. Di\'{o}si, and N. Gisin, Open System
Dynamics with Non-Markovian Quantum Trajectories, Phys. Rev. Lett. \textbf{82%
}, 1801 (1999).

\bibitem{yuyu} T. Yu, Non-Markovian quantum trajectories versus master
equations: finite temperature heat bath, Phys. Rev. A \textbf{69}, 062107
(2004).

\bibitem{jing} J. Jing and T. Yu, Non-Markovian Relaxation of a Three-Level
System: Quantum Trajectory Approach, Phys. Rev. Lett. \textbf{105}, 240403
(2010); J. Jing, X. Zhao, J. Q. You, and T. Yu, Time-local
quantum-state-diffusion equation for multilevel quantum systems, Phys. Rev.
A \textbf{85}, 042106 (2012); Y. Chen, J. Q. You, and T. Yu, Exact
non-Markovian master equations for multiple qubit systems: quantum
trajectory approach, Phys. Rev. A \textbf{90}, 052104 (2014).

\bibitem{eplYu} J. Jing and T. Yu, Stochastic Schr\"{o}dinger equation for a
non-Markovian dissipative Qubit-Qutrit system, Euro Phys. Lett. \textbf{96},
44001, (2011).

\bibitem{feria} L. Ferialdi and A. Bassi, Exact Solution for a Non-Markovian
Dissipative Quantum Dynamics, Phys. Rev. Lett. \textbf{108}, 170404 (2012).

\bibitem{YuBipartite} J. Jing, R. Li, J. Q. You, and T. Yu, Nonperturbative
stochastic dynamics driven by strongly correlated colored noises, Phys. Rev.
A \textbf{91}, 022109 (2015).

\bibitem{gambetta} J. Gambetta and H. M. Wiseman, Non-Markovian stochastic
Schr\"{o}dinger equations: Generalization to real-valued noise using
quantum-measurement theory, Phys. Rev. A \textbf{66}, 012108 (2002).

\bibitem{hidden} J. Gambetta and H. M. Wiseman, Interpretation of
non-Markovian stochastic Schr\"{o}dinger equations as a hidden-variable
theory, Phys. Rev. A \textbf{68}, 062104 (2003).

\bibitem{Diosol} L. Di\'{o}si, Non-Markovian Continuous Quantum Measurement
of Retarded Observables, Phys. Rev. Lett. \textbf{100}, 080401 (2008); H. M.
Wiseman and J. M. Gambetta, Pure-State Quantum Trajectories for General
Non-Markovian Systems Do Not Exist, Phys. Rev. Lett. \textbf{101}, 140401
(2008).

\bibitem{reduction} A. Bassi and G. C. Ghirardi, Dynamical reduction models
with general Gaussian noises, Phys. Rev. A\textbf{\ 65}, 042114 (2002); A.
Bassi, Stochastic Schr\"{o}dinger equations with general complex Gaussian
noises, Phys. Rev. A \textbf{67}, 062101 (2003); S. Adler and A. Bassi,
Collapse models with non-white noises, J. Phys. A \textbf{40}, 15083 (2007);
A. Bassi and L. Ferialdi, Non-Markovian dynamics for a free quantum particle
subject to spontaneous collapse in space: general solution and main
properties, Phys. Rev. A \textbf{80}, 012116 (2009); L. Ferialdi and A.
Bassi, Dissipative collapse models with non-white noises, Phys. Rev. A 
\textbf{86}, 022108 (2012).

\bibitem{bassilo} A. Bassi and L. Ferialdi, Non-Markovian Quantum
Trajectories: An Exact Result, Phys. Rev. Lett. \textbf{103}, 050403 (2009).

\bibitem{alonsoQRT} D. Alonso and I. de Vega, Multiple-time Correlation
Function For Non-Markovian Interaction: Beyond The Quantum Regression
Theorem, Phys. Rev. Lett. \textbf{94}, 200403 (2005); D. Alonso and I. de
Vega, Hierarchy of equations of multiple-time correlation functions, Phys.
Rev. A \textbf{75}, 052108 (2007).

\bibitem{strunzBrownian} W. T. Strunz, L. Di\'{o}si, N. Gisin, and T. Yu,
Quantum Trajectories for Brownian Motion, Phys. Rev. Lett. \textbf{83}, 4909
(1999); W. T. Strunz and T. Yu, Convolutionless Non-Markovian master
equations and quantum trajectories: Brownian motion revisited, Phys. Rev. A 
\textbf{69}, 052115 (2004).

\bibitem{eisfeld} J. Roden, W. T. Strunz, and A. Eisfeld, Non-Markovian
quantum state diffusion for absorption spectra of molecular aggregates, J.
Chem. Phys. \textbf{134}, 034902 (2011); J. Roden, A. Eisfeld, W. Wolff, and
W. T. Strunz, Influence of Complex Exciton-Phonon Coupling on Optical
Absorption and energy Transfer of Quantum Aggregates, Phys. Rev. Lett. 
\textbf{103}, 058301 $(2009).$

\bibitem{suess} G. Ritschel, D. Suess, S. M\"{o}bius, W. T. Strunz, and A.
Eisfeld, Non-Markovian Quantum State Diffusion for temperature-dependent
linear spectra of light harvesting aggregates, J. Chem. Phys. \textbf{142},
034115 (2015).

\bibitem{chinos} X. Zhong and Y. Zhao, Non-Markovian stochastic Schr\"{o}%
dinger equation at finite temperatures for charge carrier dynamics in
organic crystals, J. Chem. Phys. \textbf{138}, 014111 (2013).

\bibitem{you} J. Jing, X. Zhao, J. Q. You, W. T. Strunz, and T. Yu,
Many-body quantum trajectories of non-Markovian open systems, Phys. Rev. A 
\textbf{88}, 052122 (2013).

\bibitem{DiosiPRL} L. Di\'{o}si and L. Ferialdi, General Non-Markovian
Structure of Gaussian Master and Stochastic Schr\"{o}dinger Equations, Phys.
Rev. Lett. \textbf{113}, 200403 (2014).

\bibitem{novikov} A. Novikov, Functionals and the random-force method in
turbulence theory, Soviet Phys. JETP \textbf{20,} 1290 (1965).

\bibitem{rio} A. A. Budini, Quantum systems subject to the action of
classical stochastic fields, Phys. Rev. A \textbf{64}, 052110 (2001).
\end{thebibliography}
\end{document}